\documentclass{article}
\usepackage{spconf,amsmath,graphicx,hyperref,booktabs}

\title{OCR-Enhanced Multimodal ASR Can Read While Listening}
\name{Junli Chen, Changli Tang, Yixuan Li, Guangzhi Sun, Chao Zhang}
\address{Department of Electrical Engineering, Tsinghua University,
Beijing, China}

\begin{document}
\ninept
\maketitle
\begin{abstract}
Visual information, such as subtitles in a movie, often helps automatic speech recognition.
In this paper, we propose Donut-Whisper, an audio-visual ASR model with dual encoder to leverage visual information to improve speech recognition performance in both English and Chinese. Donut-Whisper combines the advantage of the linear and the Q-Former-based modality alignment structures via a cross-attention module, generating more powerful audio-visual features. Meanwhile, we propose a lightweight knowledge distillation scheme showcasing the potential of using audio-visual models to teach audio-only models to achieve better performance. Moreover, we propose a new multilingual audio-visual speech recognition dataset based on movie clips containing both Chinese and English partitions. As a result, Donut-Whisper achieved significantly better performance on both English and Chinese partition of the dataset compared to both Donut and Whisper large V3 baselines. In particular, an absolute 5.75\% WER reduction and a 16.5\% absolute CER reduction were achieved on the English and Chinese sets respectively compared to the Whisper ASR baseline.

\end{abstract}


\section{Introduction}
Human perceives speech in a multimodal fashion, where visual information often provides important clues that are crucial to correct recognition of the speech content. Optical characters, such as the subtitles in a movie or the content on a slide, is one of the most direct source of visual clue that helps speech perception. However, recent advancements in end-to-end ASR systems \cite{whisper2023} focuses on audio inputs alone, which may struggle when the audio signal is noisy or contain out-of-domain words. Visual information has been studied whereas research either focus on lip movements, or extracting out-of-domain words to construct the contextual biasing list which is not end-to-end differentiable \cite{sun2022treeconstrainedpointergeneratorgraph}. On the other hand, the rapid development in computer vision, especially optical character recognition (OCR), yields powerful visual encoders such as Donut \cite{donut2022} that performs robustly under real-world scenarios.

Targeting at videos with embedded subtitles, this paper proposes Donut-Whisper, an end-to-end audio-visual speech recognition model combining the pretrained ASR and OCR encoder via an effective fusion architecture to achieve OCR-enhanced ASR. Specifically, we leverage the success of both worlds and employ the Whisper-large-V3 speech encoder \cite{whisper2023} and the Donut OCR encoder \cite{donut2022} in a single unified structure. To effectively combine the output of both encoders, inspired by the effective modality aligners in multimodal large language models (MLLMs) \cite{llava,salmonn,videosalmonn}, linear and Q-Former structures are used to project the Donut and Whisper encoder outputs respectively to a representation space to be aligned with the decoder input space. Instead of directly sending these features to the decoder, we adopt a cross-attention structure to further mix the audio and visual features for more thorough interaction. The dual encoder design, together with the fusion architecture, enables the model to read while listening, yielding better ASR performance.

To demonstrate the effectiveness of Donut-Whisper, we curate a multilingual audio-visual speech recognition dataset featuring both English and Chinese movie dialogues. The dataset contains 33 hours of video for English and 57 hours for Chinese. As a result, Donut-Whisper achieved 5.6\% absolute WER reduction on English videos and a 4.0\% CER reduction on Chinese videos the reserved test sets compared to the strong Donut baseline which is also finetuned on the same dataset. Furthermore, we discover that distilling Donut-Whisper to an audio-only model yields better performance than directly training on the text transcriptions.

The rest of this paper is organized as follows. Section \ref{sec:rel} reviews the related literature. Section \ref{sec:method} introduces the Donut-Whisper model in detail. Section \ref{sec:exp} and \ref{sec:result} describe the experimental design and results, followed by conclusino in Section \ref{sec:conclusion}. 

\section{Related Work}
\label{sec:rel}

There are two mainstream approaches in audio-visual speech recognition (AVSR). One focuses on visual features related to speech, such as lips and facial movements, and the other leverages visual text information. We review both streams as follows.

\subsection{Audio–Visual Models with Lip or Facial Cues}
Early AVSR models used the speaker’s lip motions to supplement audio. For example, LipNet applied 3D CNNs and an RNN to video frames, achieving sentence-level lipreading entirely end-to-end \cite{lipnet}. Chung introduced the “Watch, Listen, Attend and Spell” (WLAS) model, which uses two LSTM encoders, one on video, one on audio, with cross-modal attention to decode text \cite{Chung_2017_CVPR}. Afouras showed that a transformer-based AVSR (“TM-CTC”) with stacked self-attention layers can also be very effective, reporting approximately 8.2\% WER on the LRS2-BBC benchmark, which is a significant improvement compared to the 10.1\% of the audio-only model \cite{afouras2018deep}.

Modern systems often use dual encoders and large pretraining. Pan train separate front ends on large unimodal corpora, e.g. a ResNet-based visual encoder MoCo v2 pretrained on ImageNet and a Wav2Vec2 audio encoder \cite{pan2022leveraging}, and fuse them in a single Transformer. Their model concatenates audio and visual embeddings and uses a hybrid CTC/attention decoder \cite{pan2022leveraging}, yielding a nice result on LRS2 by self-supervised pretraining. Similarly, Shi introduced AV-HuBERT, a masked multimodal BERT that “masks” the audio+video stream and predicts cluster units \cite{shi2022learningaudiovisualspeechrepresentation}. AV-HuBERT learns powerful speech representations: with only 30 hours of labels, it achieved 32.5\% WER on LRS3, and even yields a 40\% relative reduction in audio-only WER \cite{shi2022learningaudiovisualspeechrepresentation}.
These AVSR models target realistic scenarios: as audio-only ASR degrades in noise or multi-speaker settings, visual lip cues help focus on the target speaker \cite{shi2022robustselfsupervisedaudiovisualspeech}.

\subsection{Models Leveraging Visual Text}

A newer AVSR direction uses on-screen text like subtitles or slides as additional context. For example, in lecture or conference videos, slides often contain technical terms or names that a pure ASR might misrecognize \cite{wang2024slidespeech}. Recent work has constructed large “slide-enriched” corpora and shown how to feed slide text into ASR. SlideSpeech is a more than 1,000h dataset of lecture videos with synchronized slides and transcripts. Its authors propose a pipeline: detect and OCR slide text, extract keywords, and bias the ASR via a contextual language model to these terms \cite{wang2024slidespeech}.

On SlideAVSR, a similar dataset of paper talks, a baseline called DocWhisper simply concatenates OCR’d slide words with the audio features into Whisper’s encoder. This yielded a 14.3\% relative WER reduction over audio-only Whisper \cite{wang2024slideavsr}. Peng also demonstrated prompt-based use of visuals. They ran CLIP on frames to generate likely word tokens and prepended these to Whisper, improving zero-shot AVSR performance \cite{peng2023prompting}. The new M3AV dataset provides 367h of lectures with richly annotated slide text and speech, explicitly evaluating contextual ASR tasks \cite{chen2024m}. These studies suggest that extracting textual context from video can significantly enhance ASR, especially for specialized vocabulary in educational or other scenarios \cite{wang2024slidespeech,wang2024slideavsr}. Note that unlike Donut-Whisper, none of these attempts can be trained in an end-to-end fashion.

\section{Method}
\label{sec:method}
\subsection{Model Architecture}
\begin{figure*}[t]
\centering
\includegraphics[width=0.95\linewidth]{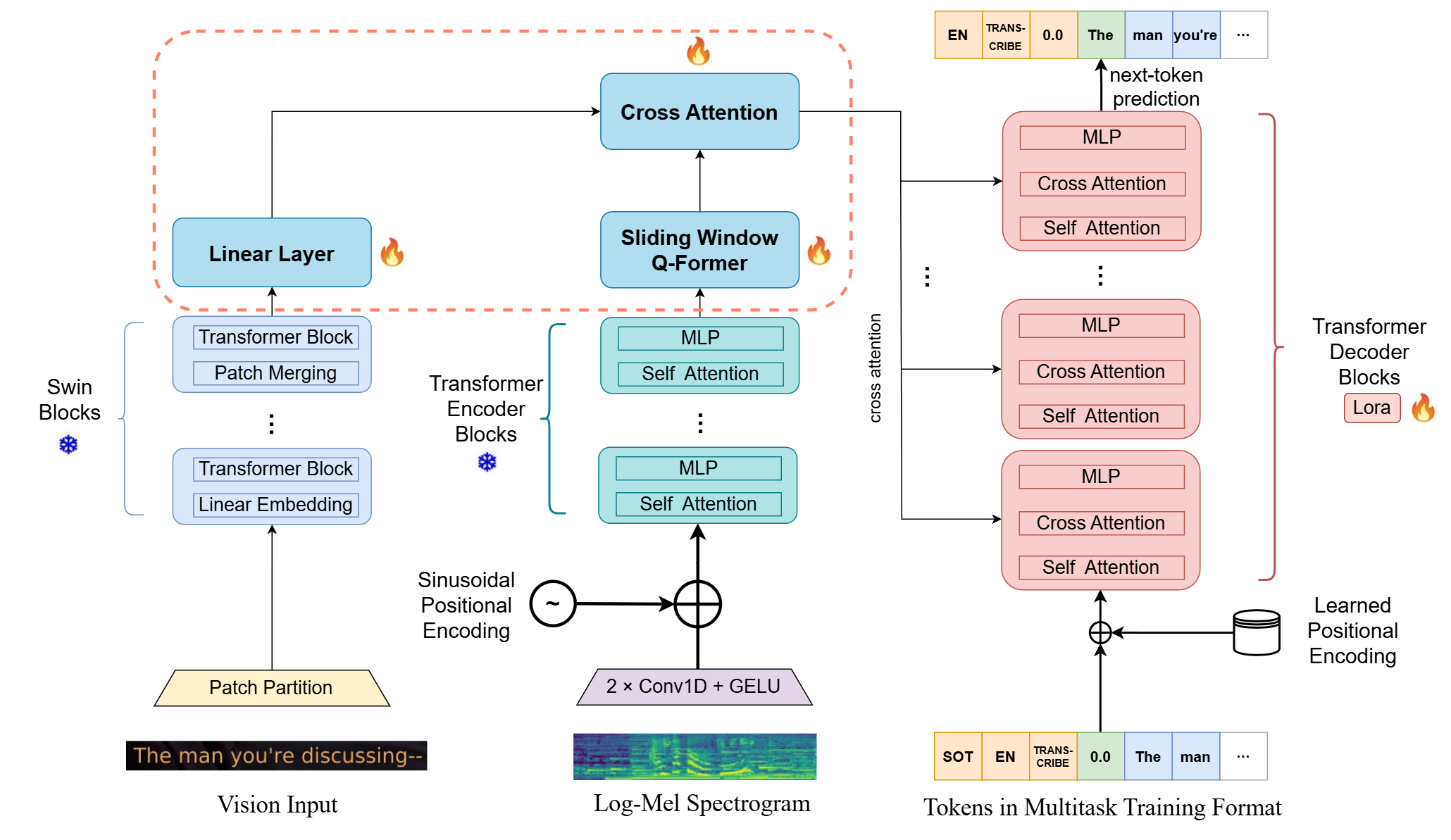}
\vspace{-0.5cm}
\caption{Architecture of the proposed Donut-Whisper model. Bottom left are the audio and visual encoders. The alignment module is shown in the dashed box. The decoder is on the right.}
\label{fig:architecture}
\end{figure*}

Figure \ref{fig:architecture} illustrates the overall architecture of Donut-Whisper. The model is composed of two encoders, one decoder, and a feature fusion module (indicated by the dashed box). 
Visual and audio encoders process visual and audio inputs in parallel.
The visual encoder is derived from the pre-trained OCR model Donut, and it takes as input key frames extracted from the video. This encoder employs a Swin Transformer to extract image features, obtaining reliable visual information \cite{donut2022} (for simplicity, only the region containing subtitles is depicted in the figure, whereas the model input is the entire video frame). The audio encoder comes from the pre-trained ASR model Whisper, with the input being the audio segment corresponding to the video. It uses multiple Transformer layers to extract features from the audio signal, converting speech into a sequence of audio features \cite{whisper2023}.

The feature fusion module consists of a linear layer, a sliding-window Q-Former \cite{salmonn,connector}, and a cross-attention layer. Considering that audio features exhibit clear temporal locality (each speech segment is most related to features in its vicinity in time), this module is designed to align and fuse the visual and auditory features. The linear layer applies a linear mapping to the visual encoder's output, projecting the visual features into the auditory feature space for subsequent fusion. The sliding-window Q-Former layer performs local aggregation on the auditory encoder's output sequence. Specifically, we divide the audio feature sequence $H^a = {h^a_1, h^a_2, \ldots, h^a_T}$ into $N$ segments of length $w$ with stride $\Delta$, as defined in Eq.~(1). 
\begin{align}
H^a_u &= [\,h^a_{(u-1)\Delta+1},\;h^a_{(u-1)\Delta+2},\;\ldots,\;h^a_{(u-1)\Delta+w}\,], \nonumber \\
\quad u &= 1, 2, \ldots, N.
\end{align}

Then, for each segment, a Q-Former module is applied to aggregate local information: the Q-Former's queries $Q$ are a set of learnable vectors, and the keys and values come from that segment's audio features. Through this local attention mechanism, we obtain an aggregated feature vector representing the local pattern of that time window. We then perform average pooling over these vectors to produce audio-enhanced queries, which are added to the learnable queries to form the module’s output of this module, as shown in Eq.~(2). The queries obtained here will be used as the queries of cross-attention in the next module.

\begin{equation}
\begin{aligned}
\tilde{h}^a_u &= \mathrm{Attention}(Q,\; H^a_u,\; H^a_u)\\
\tilde{H}^a &=[\,\tilde{h}^a_1,\tilde{h}^a_2,\ldots, \tilde{h}^a_N]\\
\tilde{h}^a &= \operatorname{mean}_{c}(\tilde{H}^a)
\end{aligned}
\end{equation}

The cross-attention layer built on the audio feature sequence extracted by the sliding-window Q-Former, then brings in the visual features (after the linear transform) as keys and values. Via a cross-modal attention mechanism, the audio features attend to the corresponding visual information, completing the fusion of audio-visual features and yielding a fused feature sequence. 

\begin{equation}
\mathrm{Attention}(Q, K, V) = \mathrm{softmax}\left(\frac{QK^T}{\sqrt{d}}\right)V.
\end{equation}

The decoder is from the pre-trained Whisper ASR model; it takes the fused feature sequence as input and uses Whisper's autoregressive decoding to generate the text output. The decoder consists of multiple Transformer layers, which progressively decode the fused features into the target text sequence \cite{whisper2023}.

\subsection{Knowledge Distillation}

We use a “small multimodal subset $\to$ unimodal adaptation” strategy without fine-tuning the multimodal teacher. From a small set of subtitle-embedded videos, the Donut-Whisper teacher (base, 0.15B params) provides (i) temperature-scaled logits ($\tau>1$) and (ii) argmax text tokens as pseudo-labels. The student is Whisper-large V3 (~1.5B, audio-only). It ingests audio, freezes the encoder, and updates the decoder via LoRA \cite{lora2022}. The distillation loss combines cross-entropy on pseudo-labels with KL on softened distributions, emphasizing CE ($\alpha>\beta$) \cite{hinton2015distilling} to follow the main output while KL transfers “dark knowledge” among secondary candidates, reducing overconfidence and aiding generalization:

\begin{equation}
L_{\text{distill}} = \alpha\,L_{\text{CE}} + \beta\,L_{\text{KL}},
\end{equation}
This prior from the multimodal teacher improves auto-labeling and final recognition even with a small distillation subset.

\section{Experiments}
\label{sec:exp}
\subsection{Dataset}
We constructed a audio-visual speech recognition dataset featuring movie clips with subtitles. The dataset contains 57 hours of videos in Chinese and 33 hours in English. Each video is annotated with the corresponding transcript text and timestamp annotations. The Chinese dataset comprises 57 hours from 244 YouTube videos, with subtitles that are provided by creator and include reference English translation, totaling approximately 103000 annotations. The English dataset comprises 37 hours from 46 TV series, with human annotated subtitles and accompanied by reference Chinese translations, totaling 42000 annotations.

To incorporate subtitle information in the visual modality, we embedded the subtitle text into the original video frames, ensuring that the model's visual input contains text aligned with the audio. For each line of subtitles with time interval $[t_s, t_e]$, we extract the middle frame of that interval as the image input, and clip the corresponding audio segment as the auditory input, thereby guaranteeing strict temporal alignment between the two modalities. When splitting into training, validation, and test sets, we ensured that no segments from the same source video appear across different sets, to avoid any cross-set information leakage. In text preprocessing, we uniformly normalized case, numbers, and Chinese/English punctuation to ensure fair evaluation across models and languages (using CER for Chinese and WER for English).

\subsection{Baselines and Metrics}
We selected Donut-base (where the visual encoder comes from) and Whisper-large V3 (the largest Whisper model) as strong unimodal baselines, and evaluated them on the above Chinese and English data to obtain baseline performance. Our multimodal models include Donut-Whisper-base and Donut-Whisper-small depending on whether we use Whisper base or small decoder, and both are using Donut-base as the visual encoder. Evaluation is conducted using Chinese character error rate (CER) and English word error rate (WER) as metrics. For evaluation, we keep the model outputs consistent with subtitles for language tags and special symbol handling to ensure comparable treatment of punctuation and cases, etc.

\subsection{Comparative Experiments}
Based on the fusion architecture, we conducted comparative experiments to assess how different fusion designs affect information-extraction capability. In the first configuration, after aligning feature dimensions, we directly concatenate the output tokens from the visual and audio encoders and pass the concatenated sequence through a linear layer before feeding it to the decoder. This keeps the number of tokens seen by the decoder unchanged, the total equals the sum of tokens from the two encoders. In the second configuration, we retain the same concatenation but replace the linear layer with a Q-Former, using the concatenated sequence as keys and values. The third configuration adopts the sliding-window Q-Former with cross-attention described earlier. For this third design, we also vary the sliding-window size to compare performance under different window settings.
\begin{figure}[t]
  \centering
  \includegraphics[width=\linewidth]{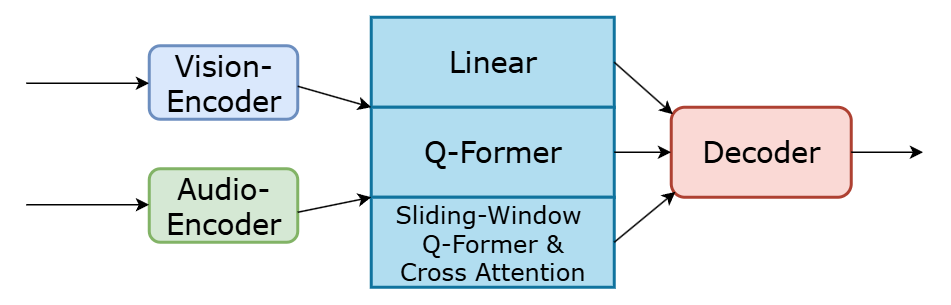}
  \caption{Architecture for alternative fusion structures under study in this paper, including a linear projection layer, a Q-Former and the proposed fusion module in Donut-Whisper.}
  \label{fig:Fusion types}
  \end{figure}

\subsection{Training Details}
When training the Donut-Whisper multimodal model, we froze the pre-trained visual and auditory encoder parameters, and only trained the feature fusion module and the Whisper decoder (using LoRA for efficient parameter updates). We used the AdamW optimizer with no weight decay \cite{loshchilov2017decoupled}, an initial learning rate of $1\times 10^{-4}$, a 3\% warm-up, and a cosine annealing scheduler. Mixed precision training was employed, and gradient clipping was applied to stabilize training.

In the knowledge distillation stage, we ensured the teacher and student models use an identical vocabulary and special control tokens to avoid token alignment ambiguities during distillation. 

\section{Results and Analysis}
\label{sec:result}
\subsection{Main Results}

Donut–Whisper (base and small) consistently outperforms the corresponding fine-tuned unimodal baselines (Donut-base and Whisper-large v3) on our Chinese–English test set (Table \ref{tab:main-results}). For Chinese, the CER reduction mainly comes from correcting homophone confusions, classifier misuse, and errors around subtitle line breaks; for English, WER improvements are most evident under high speaking rate, mild noise/echo, and short-term speaker switches. On the Chinese dataset (78,494 characters), ASR yields 14,595 errors and OCR yields 6,239, among which Donut–Whisper corrects 84.4\% (12316) and 67.9\% (4236) respectively. We attribute these gains to two factors: on the one hand, visual text provides within-line alignment and character-shape priors that help revise both ASR and OCR predictions; on the other hand, the sliding-window Q-Former localizes audio representations temporally before decoding, mitigating cross-frame attention diffusion.

\begin{table}[ht]
  \centering
  \resizebox{\columnwidth}{!}
{
  \begin{tabular}{lccc}
  \toprule
  \textbf{Model} & \textbf{Parameters} & \textbf{WER \%} & \textbf{CER \%} \\
  \midrule
  Donut base                 & 0.2B  & 9.93 & 8.7  \\
  Whisper large V3           & 1.5B   & 10.08 & 21.2  \\
  Donut-Whisper base (Ours)  & 0.15B & \textbf{4.33} & 6.7  \\
  Donut-Whisper small (Ours) & 0.33B & 5.0 & \textbf{4.7}  \\
  \bottomrule
  \end{tabular}
  }

  \caption{The performance of Donut-Whisper compared with their respective unimodal baseline models on the test set. WER indicates the word error rate on the English part, and CER represents the character error rate on the Chinese part.}
  \label{tab:main-results}
  \end{table}
  
\subsection{Effect of Distillation}
The performance of the distilled model is shown in Table \ref{tab:distillation}. Without any in-domain fine-tuning of the teacher model, using only a small portion of the data with embedded subtitles to generate pseudo-labels (for the CE loss) and soft distributions (for the KL loss) still significantly reduced the student model's error rates (for instance, the English WER saw a notable decrease). The relatively large cross-entropy loss weight helped the student model first align to the teacher's primary output path, accelerating convergence, while the temperature-scaled KL divergence provided the relative relationships among the teacher's secondary outputs, alleviating the over-confidence issue from using only hard labels. Overall, this distillation process not only improved final recognition accuracy but also markedly enhanced the automatic labeling quality on the full unlabeled dataset, paving the way for more reliable pseudo-labels in subsequent iterative self-training.

\begin{table}[ht]
  \centering
  \begin{tabular}{lcc}
  \toprule
  \textbf{Model} & \textbf{Parameters} & \textbf{WER \%} \\
  \midrule
  Donut-Whisper base (Teacher)   & 0.15B & 4.33 \\
  \midrule
  Whisper large V3 (No Distillation) & 1.5B & 10.08 \\
  Whisper large V3 (With Distillation) & 1.5B & \textbf{9.86} \\
  \bottomrule
  \end{tabular}

  \caption{The performance of the teacher model (Donut-Whisper) and the student model (Whisper Large V3) before and after knowledge distillation on the English part.}
  \label{tab:distillation}
  \end{table}

\subsection{Comparative Experiments}

We compared three fusion schemes in the Donut-Whisper model: Linear, Q-Former, and Sliding-Window Q-Former with Cross-Attention, and report the detailed results in Table \ref{tab:comparative1}. The third design clearly outperforms the first two, indicating that a sliding-window mechanism is more effective for audio signals with local temporal structure, yielding lower error rates than the other fusion strategies. Moreover, within the third design, in our experiments the optimal window size is 64, outperforming 32, 128 and 256. We attribute this to a trade-off: too long windows introduce redundant information, while too short windows fragment speech information. Regarding latency, window-size has minimal effect because the fusion module accounts for only about 5\% of inference time.

\begin{table}[ht]
  \centering
  \begin{tabular}{lcc}
  \toprule
  \textbf{Model} & \textbf{Parameters} & \textbf{CER \%} \\
  \midrule
  DW (Linear)           & 0.32B & 10.8 \\
  DW (Q-Former)         & 0.33B & 9.6  \\
  DW (Sliding-Window Q-Former 32) & 0.33B & 5.2 \\
  DW (Sliding-Window Q-Former 64) & 0.33B &  \textbf{4.7} \\
  DW (Sliding-Window Q-Former 128) & 0.33B & 6.1  \\
  DW (Sliding-Window Q-Former 256) & 0.33B & 9.0 \\
  Donut-base (baseline) & 0.2B & 8.7  \\
  Whisper large V3 (baseline) & 1.5B & 21.2  \\
  \bottomrule
  \end{tabular}

  \caption{The comparison of the performance of different fusion structures and the results under different window sizes for the sliding-window Q-Former. DW denotes our Donut-Whisper model.}
  \label{tab:comparative1}
  \end{table}

In addition, we found that even inserting only the sliding-window Q-Former module between the encoder and decoder could bring a considerable improvement in recognition performance. This indicates that the sliding-window Q-Former, as a pluggable intermediate layer, has independent value even in a unimodal ASR context. The results are presented in Table \ref{tab:comparative 2}.

\begin{table}[ht]
  \centering
  \begin{tabular}{lcc}
  \toprule
  \textbf{Model} & \textbf{Parameters} & \textbf{CER \%} \\
  \midrule
  Whisper large V3                  & 1.5B & 21.2 \\
  Whisper large V3 (Sliding-window) & 1.5B & \textbf{16.6}  \\
  \bottomrule
  \end{tabular}
  \caption{The performance comparison of Whisper before and after incorporating the sliding-window Q-Former.}
  \label{tab:comparative 2}
  \end{table}

\section{Conclusion}
\label{sec:conclusion}

This paper proposes Donut-Whisper, an end-to-end audio-visual speech recognition model targeting at understanding scenarios with optical textual clues. Donut-Whisper is not only an effective new multimodal ASR model, but also a “teacher engine” that can be used for unsupervised distillation. Future work will focus on (1) performing iterative self-training on larger-scale unlabeled corpora combined with confidence-based filtering strategies to further exploit the potential of multimodal models in low-resource settings; and (2) exploring adaptive sliding-window sizes and learnable query vector quantities, to extend the model's support for more fine-grained timestamp alignment.

\bibliographystyle{IEEEbib}
\bibliography{strings,refs}

\end{document}